**Angular dependence of the magnetization reversal in exchange biased Fe/MnF$_2$**

Elke Arenholz

Advanced Light Source, Lawrence Berkeley National Laboratory, Berkeley, CA 94720

Kai Liu

Department of Physics, University of California, Davis, CA 95616

Abstract

A detailed study of exchange-biased Fe/MnF$_2$ bilayers using magneto-optical Kerr Effect shows that the magnetization reversal occurs almost fully through domain wall nucleation and propagation for external fields parallel to the exchange bias direction. For finite angles $\phi$ between bias and external field the magnetization is aligned perpendicular to the field cooling direction for a limited field range for decreasing fields. For external fields perpendicular to the bias direction the magnetization aligns with the field cooling direction for descending and ascending fields before fully reversing. The field range for which the magnetization is close to perpendicular to the external field can be estimated using a simple effective field model.

The magnetization reversal in a ferromagnet (FM) / antiferromagnet (AFM) bilayer can be significantly altered upon cooling in an external magnetic field below the Néel temperature of the AFM [1]. In such "exchange biased" bilayers, the coercivity is enhanced [2], the loop shifts away from zero field to positive [3, 4] or negative field values [1, 2], vertical loop shifts have been observed [5] as well as asymmetrically shaped loops [6]. However, only very few studies detailed direct observations of asymmetric reversal processes, that is domain nucleation and growth for one field branch, e.g. ascending fields, and magnetization rotation for the opposite field branch, e.g. descending fields [7-11]. For example, X. Portier *et al*. [7] studied NiFe layers coupled to a range of antiferromagnetic films ($Fe_{50}Mn_{50}$, $Ir_{50}Mn_{50}$, and $Ni_{25}Co_{75}O$). For increasing fields the reversal is dominated by domain nucleation and growth, but for decreasing fields the reversal occurs by moment rotation followed by domain nucleation and growth. Bomqvist *et al*. [8] used Photoemission Electron Microscopy to show that the domain configuration at zero field in exchange biased Fe/MnPd bilayers is very different for descending fields (single domain) as opposed to ascending fields (domains with the magnetization mostly along to the bias direction). M. Gierlings and coworkers [9] reported a reversal through domain wall motion in the decreasing field branch and rotation in the increasing field branch for the exchange biased Co/CoO system. To date, one of the most thorough studies of asymmetric magnetization reversal has been conducted by Fitzsimmons *et al*. [10] and Leighton *et al*. [11] on $Fe/MnF_2$. They found that when the field is reduced from positive saturation, the magnetic moments align with a direction orthogonal to the original easy axis, before the magnetization is fully reversed. The field range that permits the perpendicular alignment increases with improved film quality [11]. As the field is reduced from negative saturation, the unidirectional anisotropy stimulates nucleation of domains and rotation is suppressed.

In this paper, we report a detailed angular-dependence study of the Fe/MnF$_2$ model system using magneto-optical Kerr effect (MOKE). Our results demonstrate that the magnetization reversal depends sensitively on the alignment between the external field and the bias direction (angle $\phi$). With a perfect alignment ($\phi=0$), the magnetization reversal occurs almost completely through domain wall nucleation and propagation. For $\phi > 4^o$, during the decreasing field sweep the magnetization aligns almost completely to a stable state perpendicular to the cooling field direction. With increasing $\phi$, the field range that can sustain such a perpendicular moment increases. In contrast, along the ascending field sweep the magnetization is largely independent of $\phi$. If the external field is applied perpendicular to the field cooling direction, the magnetization is oriented parallel to the bias direction, i.e. perpendicular to the external field, for a limited field range in both loop branches. The field range for which the perpendicular orientation is possible can be calculated using a simple effective field model.

Details of the sample preparation and characterization were given in previous publications [3, 4]. The layer structure is MgO(100) / 250 Å ZnF$_2$(110) / 500 Å MnF$_2$(110) /120 Å Fe/50 Å Al. The MnF$_2$ layer grows as twinned quasi-epitaxial thin film. One AF crystal domain is oriented with [1-10] MnF$_2$ || [110] MgO and the other domain is oriented with [001] MnF$_2$ || [110] MgO. The Fe overlayer is polycrystalline. The sample was cooled from 150 K through the Néel temperature of the MnF$_2$ layer, T$_N$

= 67 K, in a field, $H_{fc}$ = 1000 Oe. The cooling field was applied along the direction bisecting the [001] axes of the two AF domains, i.e. the easy anisotropy axes of the AF layer. This geometry causes a frustration of the coupling between the individual AF twins and the ferromagnetic overlayer and leads to an effective "45° coupling," resulting in two easy axes for the ferromagnet bisecting the easy axes (the [001] axes) of the two AF domains [11].

The MOKE experiments were performed with *p*-polarized light in the longitudinal geometry using an eight pole magnet allowing us to apply magnetic fields up to 0.8 T in any direction to the sample surface [12]. The MOKE set up is very similar to the one described by Ohldag *et al*. [13]. The angle of light incidence was 45° to the sample surface. *M-H* MOKE loops were measured at T = 15 K with the magnetic field oriented in the plane of the sample at angles $\phi = \phi_0 + \Delta\phi$ with $\phi_0 = 0°$, 90° and $-35° \leq \Delta\phi \leq 35°$ to the plane of incidence of the laser light. The bias direction of the sample was oriented either parallel or perpendicular to the plane of light incidence. This produces $M_{//}$-*H* and $M_\perp$-*H* MOKE loops which can be used to determine the orientation and relative strength of the magnetization relative to the cooling field direction as sampled over the area of the laser beam [14].

Figure 1 shows the field dependence of the magnetization components parallel and perpendicular to the cooling field direction for $\phi_0 = 0°$ and different angles $\Delta\phi$ between the external field and the bias direction. For $\phi = \Delta\phi = 0°$ and descending fields no perpendicular component is observed, i.e. the reversal mechanism occurs through domain nucleation and motion. For ascending fields a small perpendicular component is found. Its magnitude is almost independent of $\phi$. For an angle $\phi = \Delta\phi = 5°$ (Figure 1(b)) the Fe moments are oriented perpendicular to the external field for descending fields between $H_1 = -125$ Oe and $H_2 = -150$ Oe. With increasing angle $\Delta\phi$ the field range $|H_2 - H_1|$ increases. The magnetization always aligns with $H^{\perp}_{ext}$, i.e. the perpendicular magnetization component reverses it sign with $\Delta\phi$.

The magnetization reversal for external fields applied close to perpendicular to the field cooling direction, i.e. $\phi_0 = 90°$, is shown in Figure 2. Independently of $\Delta\phi$, the magnetization aligns with the easy direction given by the bias for ascending as well as descending field. For increasing $\Delta\phi$, the field range for which the Fe moments are aligned with the bias direction decreases with descending fields and increases for ascending fields.

In a simple model, similar to the one used by Beckmann *et al*. [15], the effective field acting on the ferromagnetic layer can be viewed as a superposition of three components: the exchange field of the antiferromagnet $H_X$ (>0) aligned with the bias direction, the anisotropy field $H_A$ (>0) aligned with the easy axes of the ferromagnet as well as the external field $H_{ext}$. For $\phi_0 = 0$, the external field can be decomposed into two orthogonal components $H^{\parallel}_{ext} = H_{ext} \cos\Delta\phi$ parallel to the cooling field direction and $H^{\perp}_{ext} = H_{ext} \sin\Delta\phi$ perpendicular to it. The anisotropy field $H_A$ depends on the projection of the magnetization on the easy axis and therefore points to opposite directions on opposite sides of the hysteresis loop. Acting on the ferromagnet for decreasing fields close to the onset of reversal is $H^{\parallel} = H_X + H_A + H^{\parallel}_{ext} = H_X + H_A + H_{ext}\cos\Delta\phi$ along the cooling field direction with $H_{ext}<0$, i.e. the external field $H_{ext}$ is opposite to $H_A$ and $H_X$. Only the external field component $H^{\perp}_{ext} = H_{ext}\sin|\Delta\phi|$ favors a reorientation of the Fe moments perpendicular to the bias direction. In the simplest approximation the magnetization will remain parallel to the bias direction for $|H^{\parallel}| \geq |H^{\perp}|$ and reorient to a perpendicular orientation for $|H^{\parallel}| \leq |H^{\perp}|$. Consequently, the magnetization reorientation occurs at $H_1 = -(H_A + H_X)/(\cos\Delta\phi + \sin|\Delta\phi|)$. Using the same description a transition from perpendicular to antiparallel to the field cooling direction will occur for $|H^{\parallel}| \geq |H^{\perp}|$. In this case $-H^{\parallel}_{ext} \geq H_X > 0$ is necessary to align the moments opposite to the bias direction and therefore $|H^{\parallel}| = -H_X + H_{ext}\cos\Delta\phi$. The effective field along the magnetization is $|H^{\perp}| = H_A - H_{ext}\sin|\Delta\phi|$. Therefore the magnetization realigns at $|H_2| = -(H_A + H_x)/(\cos\Delta\phi - \sin|\Delta\phi|)$.

For ascending fields $H_X$ and $H^{\parallel}_{ext}$ (> 0) are opposite to $H_A$ and consequently $H^{\parallel} = -H_A + H_X + H_{ext}\cos\Delta\phi$ and $H^{\perp} = H_{ext}\sin\Delta\phi$. A transition to a perpendicular state will occur at $H_3 = (H_A - H_x) / (\cos\Delta\phi - \sin|\Delta\phi|)$. Similarly, a transition from a perpendicular to a parallel orientation will occur at $H_4 = (H_A - H_X) / (\cos\Delta\phi - \sin|\Delta\phi|)$ which is identical to $H_3$. That the fields for reorientation to and from a perpendicular state are identical indicates that a perpendicular orientation of the magnetization to the bias direction is not a stable configuration for ascending fields. That a small perpendicular component is observed for a very limited field range is due to the fact that $H_X$ and $H_A$ show an small angular distribution in any real sample Indeed in the Fe/MnF$_2$ bilayer the magnetization reverses for $H_{ext}$ around 30 Oe for ascending fields, i.e. fields $H_X$ and $H_A$ acting on the magnetization are aligned but opposite. As discussed by Beckmann *et al.*

[15] this favors the nucleation of domains magnetized parallel to the external field and bias direction and magnetization reversal is achieved though domain wall motion.

A very similar situation occurs for descending fields and $\phi = 0°$. The effective field $H^{\parallel} = H_{ext} - H_X - H_A$ is aligned with the magnetization but opposite to it leading to a reversal through domain nucleation and growth. We expect this behavior for any system with a strong unidirectional or uniaxial anisptropy.

We can extend the effect field description to $\phi_0 = 90°$ and obtain the following expressions: $H_1 = -(H_A - H_X)/(\cos\Delta\phi - \sin|\Delta\phi|)$, $H_2 = -(H_A + H_X)/(\cos\Delta\phi + \sin|\Delta\phi|)$, $H_3 = (H_A - H_X)/(\cos\Delta\phi + \sin|\Delta\phi|)$, $H_4 = (H_A - H_X)/(\cos\Delta\phi - \sin|\Delta\phi|)$. For $\phi = 0°$, we find $H_1 = H_2 = -(H_X + H_A)$ and $H_3 = H_4 = H_A - H_X$ and consequently $H_X = 54$ Oe and $H_A = 87$ Oe for the exchange biased Fe/MnF$_2$ system under consideration here (see Figure 1(a)). Using the experimental values we can determine the $\Delta\phi$ dependence of $H_i$, $i = 1, \ldots, 4$ for $\phi_0 = 0°$ and $90°$. The results are shown in Figure 3 in comparison with experimental results. The simple effective field describes the experimental results very well while not containing a single free parameter. It accounts for the dominant features of the $\Delta\phi$ dependence of $H_i$, $i=1,\ldots, 4$.

Leighton *et al*. [11] reported that the field range for which a perpendicular orientation of the magnetization to the bias direction is observed increases with improved film quality (for constant $\phi$). This can be explained using the effective field description by considering that the anisotropy field $H_A$ is increasing with improved film crystallinity.

Consequently, $H_X + H_A$ and with that $|H_2 - H_1|$ increases for better quality films (for fixed $\phi$). Additional experiments on samples of different crystalline quality are needed to confirm this interpretation.

In summary, applying an external field as angles $0° < |\Delta\phi| \leq 35°$ to the bias direction results in a reorientation of the Fe moments perpendicular to the field cooling direction for descending fields and magnetization reversal largely through domain nucleation and propagation for ascending fields. For fields near $\phi_0 = 90°$ the magnetization aligns with to the easy direction defined by the cooling field for ascending as well as descending fields. The field range for which the magnetization is oriented perpendicular to the external field can be estimated using a simple effective field model, taking external field, exchange bias field, and anisotropy field of the ferromagnetic layer into account.

A detailed theory of magnetic exchange bias obviously needs to correlate the microscopic structure of a system with the magnetization reversal process. However, a simple and intuitive description as achieved with the effective field model is very valuable for a basic understanding of the main characteristics of the effect.

We thank Chris Leighton and Ivan K. Schuller for contributing the samples and helpful discussions. Work at UCD supported by UC-CLE. The Advanced Light Source is supported by the Director, Office of Basic Energy Sciences, of the U.S. Department of Energy.

**References**

[1]   W. H. Meiklejohn and C. P. Bean, Phys. Rev. **102**, 1413 (1956).

[2]   For a review see for example: J. Nogués and I.K. Schuller, J. Magn. Magn. Mater. **192**, 203 (1999).

[3]   C. Leighton, J. Nogués, H. Suhl, and I.K. Schuller, Phys. Rev. B **60**, 12837 (1999).

[4]   C. Leighton, J. Nogués, B.J. Jonsson-Åkerman, and I. K. Schuller, Phys. Rev. Lett. **84**, 3466 (2000).

[5]   J. Nogués, C. Leighton, and I. K. Schuller, Phys. Rev. B **61**, 1315 (2000).

[6]   T. Ambrose and C.L. Chien, J. Appl. Phys. **83**, 7222 (1998).

[7]   X. Portier, A. K. Petford-Long, A. de Morais, N. W. Owen, H. Laidler, and K. O'Grady, J. Appl. Phys. **87**, 65412 (2000).

[8]   P. Blomqvist, Kannan M. Krishnan, and H. Ohldag, Phys. Rev. Lett. **94,** 107203 (2005).

[9]   M. Gierlings, M.J. Prandolini, H. Fritzsche, M. Gruyters, and D. Riegel, Phys. Rev. B **65**, 092407 (2002).

[10]  M.R. Fitzsimmons, P.C. Yashar, C. Leighton, J. Nogués, J. Dura, C.F. Majkrzak, and I.K. Schuller, Phys. Rev. Lett. **84**, 3986 (2000).

[11]  C. Leighton, M.R. Fitzsimmons, P. Yashar, A. Hoffmann,  J.Nogués, J. Dura, C.F. Majkrzak, and I.K. Schuller, Phys. Rev. Lett. **86**, 4394 (2001).

[12]  E. Arenholz, S.O. Prestemon, AIP Conference Proceedings **705**, 1170 (2004).

[13]  H. Ohldag, N. B. Weber, F. U. Hillebrecht, and E. Kisker, J. Appl. Phy**s. 91**, 2228 (2002).

[14] C. Daboo, R.J. Hicken, E.Gu, M. Gester, S.J. Gray, D.E.P. Eley, E. Ahmad, J.A.C. Bland, R. Ploessl, and J.N. Chapman, Phys. Rev. B **51**, 15964 (1995).

[15] B. Beckmann, U. Nowak, and K. D. Usadel, Phys. Rev. Lett. **91,** 187201 (2003).

**Figure Captions**

**Figure 1:**

MOKE loops for external fields (black arrow) at (a) $\Delta\phi = 0°$ and (b) $\Delta\phi = 5°$ with respect to the field cooling direction (white arrow). For $\Delta\phi = 0°$ the hysteresis loop is close to square. For $\Delta\phi = 5°$ the loop show a plateau for decreasing fields for which the Fe moments are oriented perpendicular to the bias direction.

**Figure 2**

Magnetization reversal for external fields (black arrow) applied close to perpendicular to the field cooling direction (white arrow), i.e. $\phi_0 = 90°$ and (a) $\Delta\phi = 0°$, (b) $\Delta\phi = 6°$. The magnetization aligns with the easy direction given by the bias for ascending as well as descending field.

**Figure 3**

Comparison of experimental values (symbols) with the effective field model (lines) for external fields (a) near the bias direction, i.e. $\phi_0 = 0°$, and (b) close to perpendicular to the bias direction, i.e. $\phi_0 = 90°$. The field values obtained from loops measured in parallel (perpendicular) geometry, i.e. with external fields parallel (perpendicular) to the plane of light incidence, are shown by solid (open) symbols.

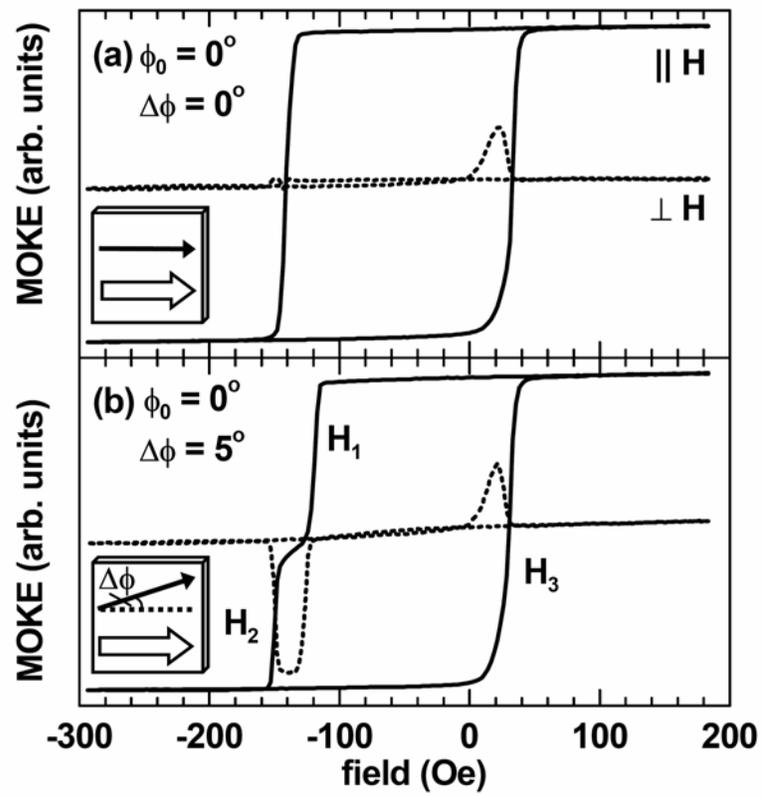

Figure 1

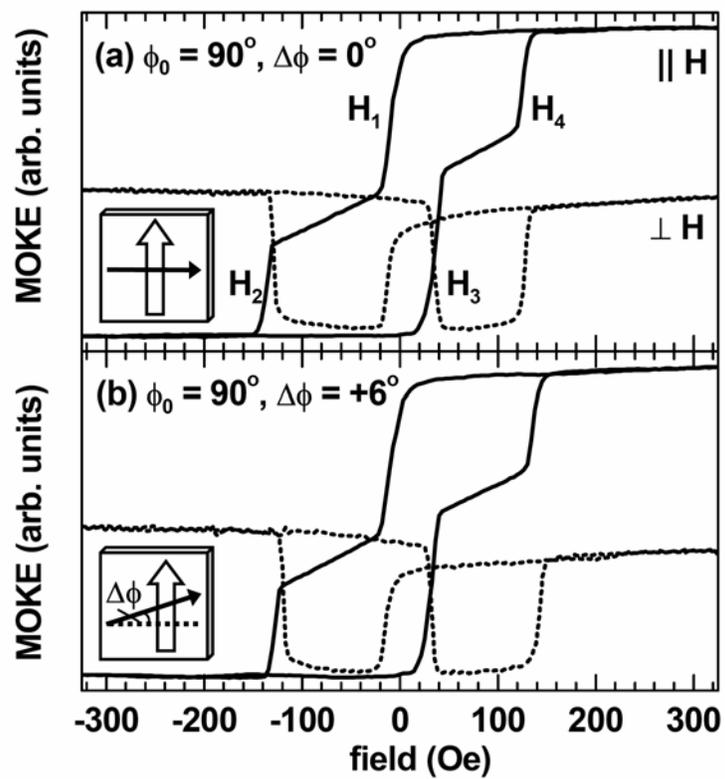

Figure 2
13

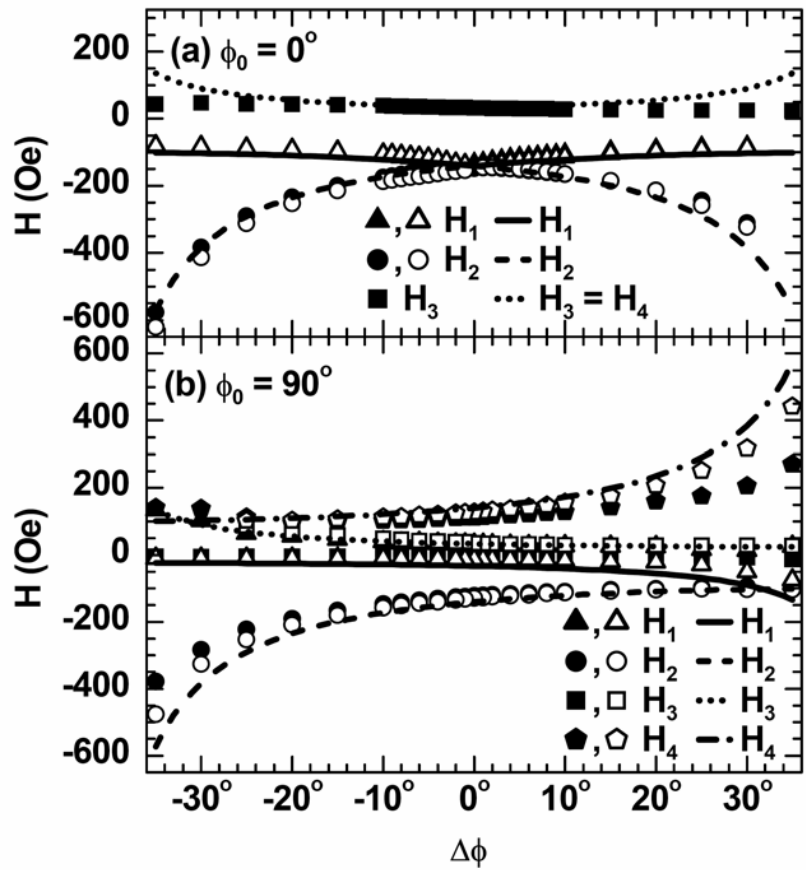

Figure 3